\documentclass[namedreferences]{solarphysics}
\usepackage{spr-sola-addons} 

\usepackage{graphicx}
\usepackage{rotating}
\usepackage{dcolumn}
\usepackage{color}           

\newcommand{\rsunNS}{$R_{S}$}
\newcommand{\rsun}{$R_{S}$ }
\newcommand{\mydeg}{$^{\circ}$ }
\newcommand{\mydegM}{^{\circ}}
\newcommand{\mydegNS}{$^{\circ}$}

\newcommand{\kmsNS}{km s$^{-1}$}

\newcommand{\aap}{    {\it Astron. Astrophys.}}

\newcommand{\apj}{    {\it Astrophys. J.}}
\newcommand{\apjl}{   {\it Astrophys. J. Lett.}}

\newcommand{\grl}{    {\it Geophys. Res. Lett.}}

\newcommand{\jgr}{    {\it J. Geophys. Res.}}

\newcommand{\procspie}{	{\it Proc. SPIE}}
\newcommand{\solphys}{{\it Solar Phys.}}

\chardef\us=`\_

\begin{document}
\begin{article}
	
\begin{opening}

\title{Deflection and Rotation of CMEs from Active Region 11158}

\author[addressref={aff1},corref,email={christina.d.kay@nasa.gov}]{\inits{C.}\fnm{Christina}~\lnm{Kay}\orcid{0000-0002-2827-6012}}
\author[addressref={aff1}]{\inits{N.}\fnm{Nat}~\lnm{Gopalswamy}\orcid{0000-0001-5894-9954}}
\author[addressref={aff1,aff2}]{\inits{H.}\fnm{Hong}~\lnm{Xie}\orcid{0000-0002-0058-1162}}
\author[addressref={aff1,aff2}]{\inits{S.}\fnm{Seiji}~\lnm{Yashiro}\orcid{0000-0002-6965-3785}}

\address[id=aff1]{Solar Physics Laboratory, NASA Goddard Space Flight Center, Greenbelt, MD, USA}
\address[id=aff2]{Department of Physics, The Catholic University of America, Washington DC, USA}

\runningauthor{Kay \textit{et al.}}
\runningtitle{CMEs from AR 11158}

\begin{abstract}
Between the 13 and 16 of February 2011 a series of coronal mass ejections (CMEs) erupted from multiple polarity inversion lines within active region 11158. For seven of these CMEs we use the Graduated Cylindrical Shell (GCS) flux rope model to determine the CME trajectory using both \textit{Solar Terrestrial Relations Observatory} (STEREO) extreme ultraviolet (EUV) and coronagraph images. We then use the Forecasting a CME's Altered Trajectory (ForeCAT) model for nonradial CME dynamics driven by magnetic forces, to simulate the deflection and rotation of the seven CMEs. We find good agreement between the ForeCAT results and the reconstructed CME positions and orientations. The CME deflections range in magnitude between 10\mydeg and 30\mydegNS. All CMEs deflect to the north but we find variations in the direction of the longitudinal deflection. The rotations range between 5\mydeg and 50\mydeg with both clockwise and counterclockwise rotations occurring. Three of the CMEs begin with initial positions within 2\mydeg of one another. These three CMEs all deflect primarily northward, with some minor eastward deflection, and rotate counterclockwise. Their final positions and orientations, however, respectively differ by 20\mydeg and 30\mydegNS. This variation in deflection and rotation results from differences in the CME expansion and radial propagation close to the Sun, as well as the CME mass.  Ultimately, only one of these seven CMEs yielded discernible \textit{in situ} signatures near Earth, despite the active region facing near Earth throughout the eruptions.  We suggest that the differences in the deflection and rotation of the CMEs can explain whether each CME impacted or missed the Earth.
\end{abstract}

\end{opening}

\section{Introduction}
Understanding the path that a coronal mass ejection (CME) takes as it propagates away from the Sun is essential for predicting any space weather effects it may induce at Earth or elsewhere in the heliosphere.  While current efforts have focused on predicting when a CME will impact Earth (\textit{e.g.} \citealt{May15AT}, and references within), one must first understand if a CME will impact Earth, and even which part of the CME will yield the impact.  This requires knowledge of any CME deflection - a deviation in latitude, longitude, or both, from a perfectly radial trajectory.  Additionally, CME rotations, changes in the orientation of the CME, can also have significant effects.

CME deflections have been observed since the earliest spaceborne coronagraph measurements \citep{Hil77, Mac86}.  \cite{Hil77} noted a systematic motion of CMEs toward the solar equator in the \textit{Skylab} observations.  \cite{Mac86} still found evidence of deflections in the \textit{Solar Maximum Mission} observations, however, the systematic equatorward motion no longer occured.  With the launch of the twin \textit{Solar Terrestrial Relations Observatory}  (STEREO) spacecraft, CMEs could be observed from more than a single viewpoint.  These additional perspectives, combined with stereoscopic reconstruction techniques, confirmed that deflections could occur in both latitude and longitude (\textit{e.g.} \cite{Isa13}; \cite{Lie15}).

Deflections were initially correlated with the relative positions of coronal features such as the heliospheric current sheet (HCS) and coronal holes (CHs).  The deflection motion was frequently described as toward the HCS \citep{Cre04, Kil09} or away from CHs \citep{Gop09}.  Typically, these two directions tend to be aligned as the HCS and CHs are intrinsically coupled by the solar magnetic field.  On global scales, CHs tend to be the regions of highest magnetic field strength, outside of active regions, and the HCS the region of lowest magnetic field strength.  The solar magnetic field reverses radial direction at HCS, which causes a decrease in the magnetic field strength near the HCS.  Often in magnetic field models the magnetic field becomes zero near the HCS, but while observations show a decrease in the magnetic strength near the HCS, it is still nonzero due to the tangential field components (\textit{e.g} \citealt{Gos05}).  Magnetic forces may be the mechanism responsible for CME deflections as on global scales they will produce the same general trends as seen in observations \citep{Fil01, Gop09, She11, Gui11, Kay15}.  Related to the rolling motion of eruptive prominences \citep{Pan11, Pan13}, motions in the low corona may be caused by smaller scale magnetic gradients related to the structure of active regions (ARs) or to local magnetic null points \citep{Kay15}.

Deflection toward the HCS, and the variation in its position over the solar cycle, could explain the difference in the observations of \cite{Hil77} and \cite{Mac86}, which respectively occurred in solar minimum and maximum.  During solar minimum the HCS is flat and near the equator and CHs are located near the poles, so primarily equatorward deflections should occur.  However, as the Sun approaches solar maximum the HCS becomes inclined and the CHs extend to low latitudes, so a wider range of deflections will occur.

Observations show evidence for CME rotation in the corona (\textit{e.g.} \citealt{Vou11}; \citealt{Nie12}; \citealt{Tho12}).  It is difficult to disentangle the effects of CME deflection, rotation, and expansion in the low corona \citep{Sav10, Nie13}, but CME rotation is also observed in simulations due to a variety of mechanisms \citep{Tor03, Fan04, Lyn09}.  Simulated rotations, such as these, tend to occur as a result of the kink instability and the direction of the rotation is then directly related to the handedness of the flux rope magnetic field. Better understanding of the rotation of CMEs is needed to understand the expected orientation of CMEs upon impact at Earth.

In this article we study the deflections and rotations of seven CMEs that all erupted from the same AR.  In particular, we compare the similarities and differences in their trajectories and seek explanation for the differences between their evolution.  In Section \ref{Obs} we describe the CME source, AR 11158, and our reconstruction of the CME positions from the coronagraph observations.  In Section \ref{ForeCAT} we describe ForeCAT, a model for CME deflections and rotations based upon magnetic forces, which we use to simulate each of the seven CMEs, shown in Section \ref{results}.  Finally, in Section \ref{Disc} we discuss the implications of the different CME deflections and rotations.

\section{Observations}\label{Obs}
\begin{figure}[!hbtp]
\includegraphics[width=4.5in, angle=0]{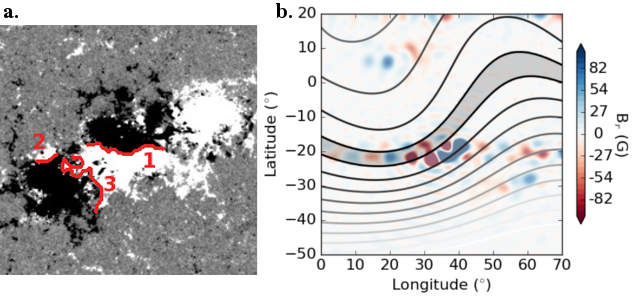}
\caption{The left panel shows an image of AR 11158 from HMI from 14 February 2011 at 03:30 UT.  The red lines indicate the three different PILs. The right panel shows the results of a potential field source surface (PFSS) magnetic field model  with color regions of the surface magnetic field near AR 11158 (at 1 \rsunNS) and line contours of the magnetic field farther out (2.5 \rsunNS) projected onto the solar surface.  The grey region indicates the location of the heliospheric current sheet, approximated by the location of the weakest magnetic field strength. Panel b shows a much larger field of view than panel a with the AR in the HMI magnetogram corresponding to the enhanced magnetic field in the center of the color regions of the surface PFSS magnetic field.}\label{fig:AR}
\end{figure}

AR 11158 was extremely active between 13 and 16 2011 February 13, and remained facing toward Earth the entire time.  During this time span 21 flares occured in AR 11158, ranging from C4.2 up to X2.1 according to the \textit{GOES} X-ray classification.  Six of the flares were M1.0 class or greater, and these larger flares occurred uniformly throughout this time span.  The flares occurred at three different polarity inversion lines (PILs) within the AR.  Figure \ref{fig:AR}a shows an image of AR 11158 from the \textit{Helioseismic and Magnetic Imager} (HMI: \citealt{HMI}) onboard the \textit{Solar Dynamics Observatory} (SDO: \citealt{SDO}) with the red lines indicating the locations of the PILs.  Two of these PILs were relatively horizontal (labeled with a 1 and a 2 in Figure \ref{fig:AR}) - a small one near the northeast of the AR and a larger one slightly northwest of the center.  The third PIL was relatively vertical, and located between and slightly south of the other two PILs.  We identify the location of each flare using SDO/\textit{Atmospheric Imaging Assembly} (AIA: \citealt{AIA}) 94 \AA{} images.  Flaring was evenly distributed between the three PILs, with multiple PILs being involved in several of the flares.  Flares from the third vertical PIL (labeled with a 3) tended to occur at later times.

Eleven of these flares had associated CMEs.  Unlike the flares, there is significant asymmetry in the temporal and spatial distribution of the CMEs.  Ten of the first eleven flares have associated CMEs, but only one of the last ten flares is accompanied by a CME.  The CMEs all erupt from one of the two horizontal PILs, no CMEs occur at the vertical PIL.  The earlier CMEs tend to come from the larger, more westward PIL (hereafter PIL 1) and the later CMEs from the smaller, more eastward PIL (hereafter PIL 2).

We determine the trajectory of these CMEs by simultaneously fitting the Graduated Cylindrical Shell (GCS) model \citep{The06} to both coronagraph views from STEREO A and B/COR1 \citep{COR1} using SolarSoft IDL procedures.  When possible we also use STEREO EUV images.  All images are first processed using the \texttt{secchi\_prep} routine.  A wireframe CME model is then fit to the dual coronagraph views.  The CME height, latitude, longitude, tilt, angular width, and a shape parameters are adjusted by hand until a visual match is obtained.  We fit each CME at multiple times throughout its evolution, obtaining the deflection and rotation of the CME from the change in its position and orientation.  Throughout this work we refer to these reconstructed positions as  the ``observations'' (as opposed to our simulated values).  We emphasize that these values are in fact free parameters of the GCS model, which is itself highly uncertain due to the complicated nature of the line-of-sight integrated coronagraph images.  However, this technique of constraining the GCS parameters using multiple simultaneous coronagraph observations is still the best available means of determining the position and orientation of a CME.

\begin{figure}[!hbtp]
\includegraphics[width=4.5in, angle=0]{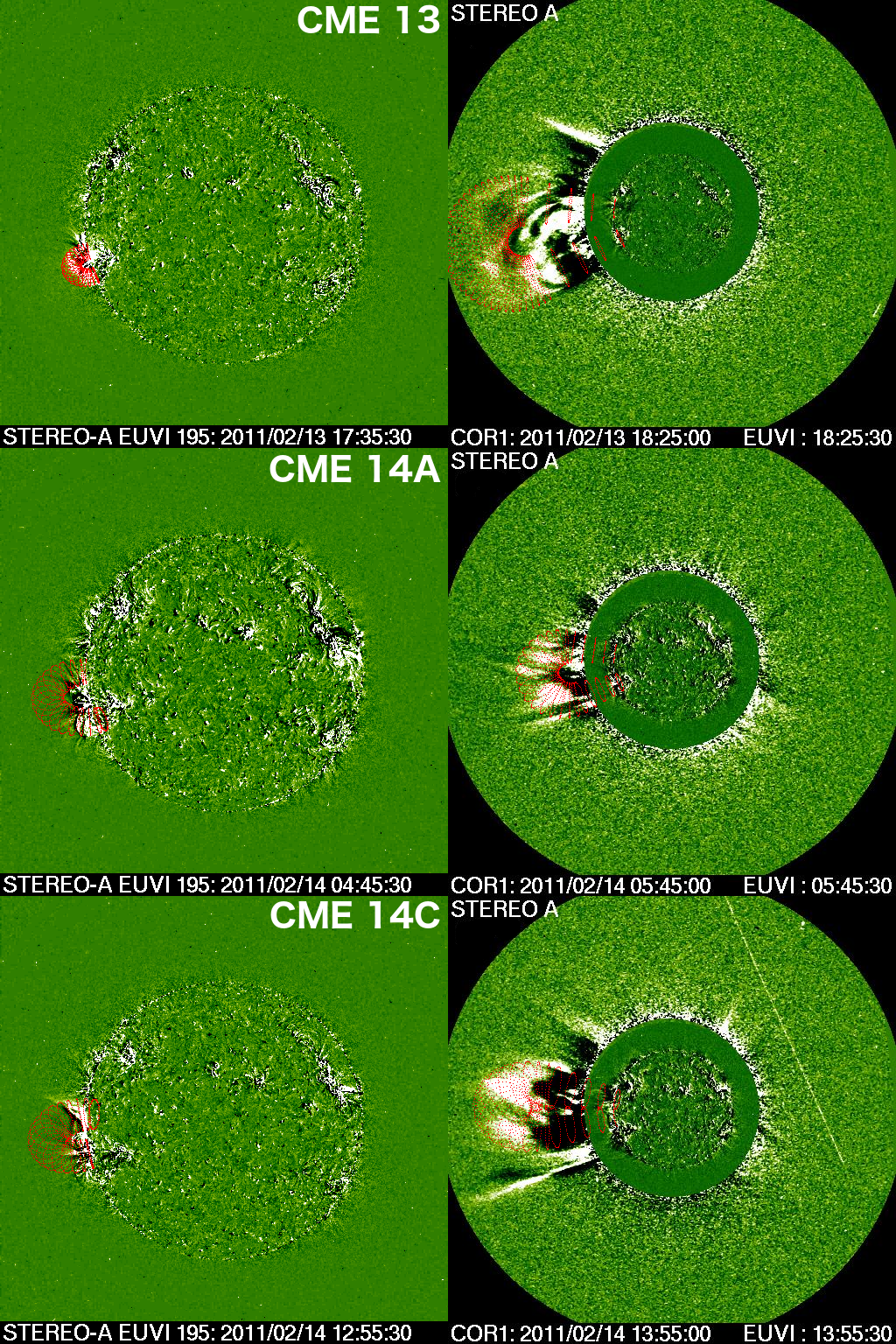}
\caption{GCS reconstructed positions for CMEs 13B (top), 14A (middle), and 14C (bottom).  The time corresponding to the first and last measured positions are shown respectively on the left and right.  The left column shows STEREO A/EUVI images and the right shows STEREO A/COR1 images with EUVI images in the center.  STEREO A was located 87\mydeg west of the Earth during the observations.  The red wireframe shows the GCS reconstruction of the CME at each height.}\label{fig:GCS}
\end{figure}

We cannot fit all the CMEs from AR 11158.  Several of the CMEs are too faint to reliably fit them using the GCS model.  One of the CMEs interacts with a preceding CME from another AR, which we do not include as simulating this collision is beyond the scope of this work.  We reproduce the trajectory for seven CMEs, one on 13, 15, and 16 February 2011, respectively, and four on 14 February 2011.  We refer to the CMEs by their day of eruption and use suffixes of A through D to differentiate between the four occurring on 14 February, assigned in chronological order.  Note that additional CMEs occurred on 14 February in between our reconstructed CMEs, but we do not consider them in our labelling scheme.  Table 1 lists the seven CMEs, the time of our first measured position, and the PIL from which they erupted.  Figure \ref{fig:GCS} shows the GCS model as a green wireframe on top of the STEREO A image for three of the CMEs (CME 13, 14A, and 14C, which have the largest deflections and rotations).  The left column shows the closest measured CME position (from STEREO/\textit{Extreme Ultraviolet Imager} (EUVI)) and the right column shows the farthest position (from STEREO/COR1).  Often the CME is difficult to resolve in the EUV but an estimate of its location can be determined from the flare brightening and the known AR PIL location.  The reconstructed CME latitude, longitude, and tilt (measured clockwise with respect to the solar equator) versus distance for all the CMEs are shown in Figures \ref{fig:CPAs1} and \ref{fig:CPAs2}.

We assume error bars of 5\mydeg for the reconstructed latitude, and 10\mydeg for the reconstructed longitude and tilt, standard values for visual GCS fits \citep{The09}.  We see that some of the CMEs show significant latitudinal deflections and rotations beyond the magnitude of the error bars.  Due to the large uncertainty, however, most of the reconstructed longitudinal motion is consistent with no deflection.  In some cases (CMEs 14B, 14D, and 15) we see little change in either the reconstructed latitude or tilt, but the value differs significantly from the initial latitude and tilt expected from our knowledge of the AR, which implies that significant evolution must have occurred before our first reconstructed points.

\begin{sidewaystable}[!htp]
\begin{center}
\newcolumntype{d}[1]{D{.}{.}{#1}}
\begin{tabular}{|l|l|c|d{1}|d{1}|d{1}|d{1}|d{1}|d{1}|}
\hline
CME & Date and Time & PIL [UT] & \mathrm{Lat}_0  \; [\mydegM] & \mathrm{Lat}_f  \; [\mydegM] & \mathrm{Lon}_0 \; [\mydegM]&  \mathrm{Lon}_f \; [\mydegM] & \mathrm{Tilt}_0 \; [\mydegM] & \mathrm{Tilt}_f \; [\mydegM] \\
    \hline
13  & 13 Feb 2011 17:35 & 1 & -20.4 & -11.1 & 35.6 & 39.2 & -34.0 & -81.4 \\
14A & 14 Feb 2011  4:45 & 1 & -20.7 &  -7.1 & 37.0 & 36.0 & -25.0 & -76.5 \\
14B & 14 Feb 2011  7:15 & 2 & -19.4 &   5.0 & 30.8 & 36.3 &   0.0 & -40.9 \\
14C & 14 Feb 2011 12:55 & 2 & -18.9 &   7.8 & 31.5 & 20.5 &  30.0 &  37.5 \\ 
14D & 14 Feb 2011 19:45 & 2 & -20.6 & -11.7 & 31.5 & 31.3 &  40.0 &  69.6 \\
15  & 15 Feb 2011  1:50 & 1 & -19.0 & -12.2 & 33.5 & 35.7 &   5.0 & -12.6 \\
16  & 16 Feb 2011 14:25 & 2 & -18.9 &   4.4 & 31.5 & 22.3 &  30.0 &  37.5 \\
    \hline
\end{tabular}
\caption{Initial ($0$) and final ($f$) positions and orientation of the considered CMEs from the results of the ForeCAT simulations.  The longitudes are given as Carrington longitudes.}
\label{tab:CMEs}
\end{center}
\end{sidewaystable}

\section{ForeCAT}\label{ForeCAT}
\cite{Kay13} first introduced the model Forecasting a CME's Altered Trajectory (ForeCAT), which simulates CME deflections resulting from the magnetic forces from the background solar magnetic field.  \cite{Kay15} expanded upon ForeCAT, allowing for deflections in both latitude and longitude, and incorporated the effects of rotation due to differential deflection forces along the CME producing a torque about the CME nose.  Note that this external torque can result in either direction rotation, whereas the kink instability driven rotation from the internal CME magnetic field, which is not currently included in ForeCAT, can only yield a single direction determined by the CME handedness. ForeCAT simulates the nonradial motion of a CME from the background magnetic tension and magnetic pressure gradients.  To describe the background solar magnetic field we use our own Python implementation of the PFSS model \citep{Alt69, Sch69} computed from an HMI synoptic magnetogram and with the source surface set to the standard distance of 2.5\rsunNS. Figure \ref{fig:AR}b shows the PFSS model for Carrington rotation 2106 near AR 11158.  The color regions show the magnetic field at the surface (1 \rsunNS), which can be compared to the HMI observations in Figure \ref{fig:AR}a.  The line contours show the magnetic field strength farther out (2.5 \rsunNS) with darker lines indicating weaker magnetic field strength.  The grey-shaded region indicates the minimum magnetic field strength, which corresponds to the location of the HCS.  ForeCAT uses the direction of the magnetic field from the PFSS along with the shape and orientation of the CME to approximate the draping of the solar magnetic field about the CME.  For a more thorough description of the ForeCAT model see \cite{Kay15}.

ForeCAT simulations require the initial position and orientation of the CME, which is constrained by the observed flare location, the CME shape, which is typically unconstrained, and the CME mass, angular width, and radial velocity as a function of time or radial distance.  We use the coronal reconstructions to constrain the expansion and propagation.  The CME mass, which we treat as constant in these cases, the CME shape, and the precise value of the initial latitude, longitude and tilt are all free parameters.  The best fit between the reconstructed deflection and rotation and the ForeCAT results allows us to constrain these previously unknown parameters.

ForeCAT currently requires an empirical model of a CME radial propagation and expansion.  In this work we assume all CMEs have a three-phase radial propagation, similar to that of \cite{Zha06}.  A CME initially propagates with some minimum velocity, $v_{min}$, until it reaches a radial distance $R_1$.  Upon the CME nose reaching $R_1$ the CME begins accelerating at a linear rate until it reaches some final velocity, $v_{f}$, at $R_2$.  We define the CME expansion in terms of half of the face-on angular width (hereafter simply angular width), which is the value returned from the SolarSoft GCS procedure.  Since the GCS wireframe was fit to both STEREO spacecraft, as opposed the view from the \textit{Large Angle Spectrometric Coronagraph} (LACSO), we find a slight difference in the angular width as compared to the LASCO catalog\footnote{\url{https://cdaw.gsfc.nasa.gov/CME\_list/UNIVERSAL/2011\_02/univ2011\_02.html}}, which also quotes the full face-on angular width.
We assume the CMEs have an expansion that follows the form
\begin{equation}
\theta = \theta_0 + \theta_M (1 - \exp^{-(R-1)/R_{\theta}})
\end{equation}
where $\theta$ is the angular width of the CME, $R$ is the radial distance of the CME nose from the Sun center, and $\theta_0$, $\theta_M$, and $R_{\theta}$ are free parameters.  We also include a maximum angular width, $\theta_f$, beyond which the angular width remains constant.  We find that the deflection and rotation is not typically sensitive to this chosen $\theta_f$ since the magnetic forces have already become negligible by the distance the CME reaches $\theta_f$.  

It is often difficult to distinguish between CME deflection, rotation, and expansion in the low corona (e.g. \citealt{Nie12}), making precise observational determination of our model constraints implausible.  \cite{Kay16Obs} show that ForeCAT can be used to better understand the early evolution of deflecting and rotating CMEs as we can identify and eliminate initial parameters that do not reproduce the observed behavior at farther distances.  In this article we use the observations to constrain some parameters such as the final speed and angular width, but for the rest we must explore the parameter space and find those values that successfully reproduce the observed CME deflections and rotations.  We will explore the effects of differences in the expansion and propagation models.

\section{Results}\label{results}
\begin{figure}[!hbtp]
\includegraphics[width=4.5in, angle=0]{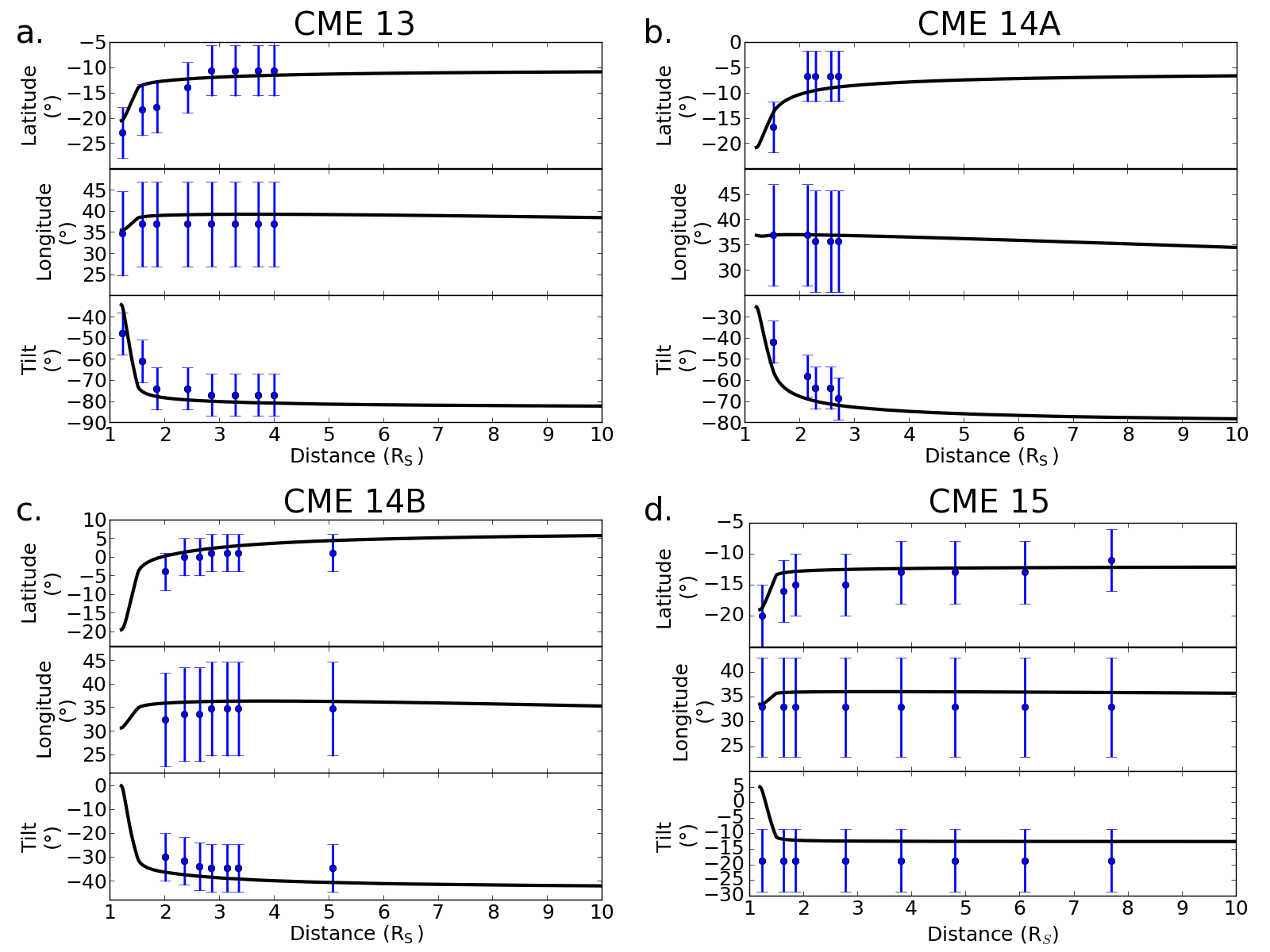}
\caption{Comparison of the positions (latitude and longitude) and orientations of the reconstructed CMEs (blue circles) with the ForeCAT results (black lines) for four of the CMEs.}\label{fig:CPAs1}
\end{figure}

\begin{figure}[!hbtp]
\includegraphics[height=5in, angle=0]{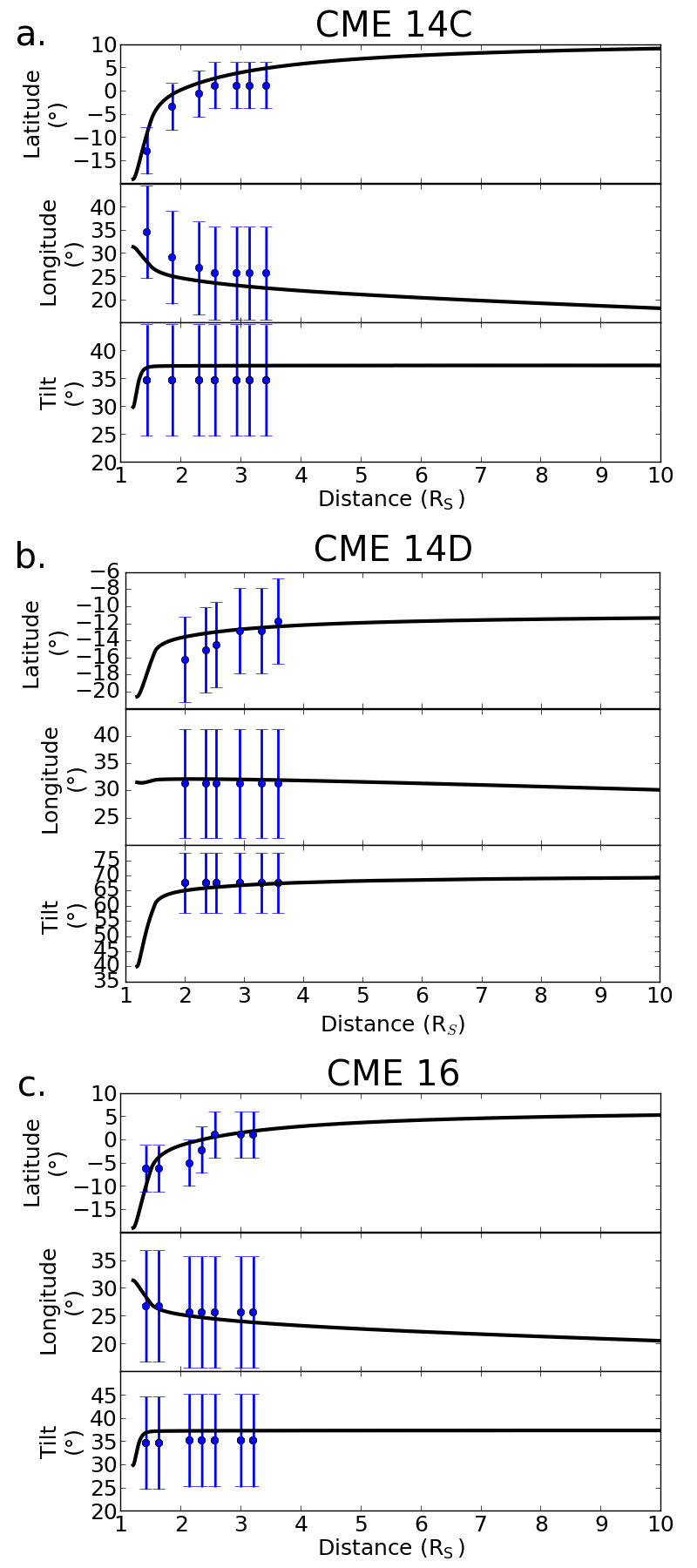}
\caption{Same as Figure \ref{fig:CPAs1} but for the other three CMEs, which all have initial positions within 2\mydeg of one another.}\label{fig:CPAs2}
\end{figure}

Using ForeCAT, we simulate each of the seven observed CMEs.  Figure \ref{fig:AR}b shows that AR 11158 is located almost directly underneath, though slightly to the south of the HCS.  We expect that deflection toward the HCS will move the CMEs northward a small amount, but the deflection could continue east or westward along the current sheet, or be influenced by the small scale gradients within the AR.  Table \ref{tab:CMEs} shows the initial and final CME positions from the ForeCAT simulations.  The black lines in Figures \ref{fig:CPAs1} and \ref{fig:CPAs2} show the ForeCAT results.  Figure \ref{fig:CPAs1} contains CMEs from both PIL 1 and 2 but the three CMEs included in Figure \ref{fig:CPAs2} are the three CMEs that erupted from PIL 2 and have initial positions within 2\mydeg of one another.  For all seven CMEs we are able to reproduce both the observed deflection and rotation within the limits of the reconstruction technique.  

We first note that, despite all originating from the same AR, the seven CMEs have very different behavior.  All seven CMEs deflect northward, but the magnitude of the latitudinal deflection varies between 8.9\mydeg and 26.7\mydegNS.  For all cases the latitudinal deflection exceeds the longitudinal deflection, which varies between 0.2\mydeg and 10\mydegNS.  Both eastward and westward deflections occur.  We also see large differences in the CME rotation with both clockwise and counterclockwise rotation occurring and magnitudes ranging between 5.7\mydeg and 51.5\mydegNS.

For all CMEs we find a rapid deflection below about 2 \rsunNS, after which the deflection either becomes negligible or proceeds at a much slower rate.  This trend has been previously noted in \cite{Kay13}, \cite{Kay15L}, and \cite{Kay15AM}. When the deflection does continue beyond 2 \rsun this motion rarely exceeds an additional deflection of 5\mydeg in latitude or longitude.  We see the same behavior for the CME rotation with the most rapid rotation occurring below 2 \rsunNS.

Both CMEs 13 and 14A erupt from PIL 1, with similar initial latitudes, but over a degree difference in their initial latitude.  Both CMEs are relatively weak, small CMEs, but CME 13 is slightly more massive (7$\times$10$^{14}$ g), faster (570 \kmsNS) and larger angular width (22\mydegNS) than CME 14A (4$\times$10$^{14}$ g, 300 \kmsNS, 17\mydegNS).  Note that the masses are determined from the best fit ForeCAT results and the speed and angular width are constrained by the observations.  Figures \ref{fig:CPAs1}a and b shows relatively similar behavior for the two CMEs.  Both deflect northward and rotate clockwise to a nearly vertical orientation, but the magnitudes of the deflection and rotation are greater for CME 13.  Neither CME exhibits a significant longitudinal deflection but the direction of the motion differs for the two cases.  CME 13 continually deflects westward until the motion becomes negligible around 2 \rsun where CME 14A oscillates from east to west back to a final eastward motion.  This shows the susceptibility of very low mass CMEs to differences in the local magnetic gradients in the low corona, however, the net longitudinal motions remain small in this case.

Figure \ref{fig:CPAs1}d shows CME 15, which is the most massive (10$^{15}$ g), fastest (1600 \kmsNS), and largest (37\mydegNS) CME considered in this article.  Compared to CMEs 13 and 14A it erupts at a slightly higher latitude and more eastward longitude along PIL 1.  CME 15 also begins with an orientation more parallel to the equator.  Despite these differences, CME 15 shows the same behavior as CME 13 and CME 14A (excluding the longitudinal oscillation), but the magnitudes of the deflection and rotation are much smaller due to the increased CME mass and speed.

CME 14B (Figure \ref{fig:CPAs1} c), which erupts from PIL 2, is a relatively low mass (5x10$^{14}$ g), slow (400 \kmsNS), and small (17\mydegNS) CME.  It deflects northward, similar to the PIL 1 CMEs, but with a significantly larger magnitude.  The westward deflection is comparable to that of CME 13.  CME 14B rotates roughly 40\mydeg clockwise, comparable but slightly smaller than the clockwise rotations of CMEs 13 and 14A.  CME 14B begins with a much flatter orientation, so it does not become nearly vertical like CMEs 13 and 14B.

CMEs 14C, 14D, and 16 (Figures \ref{fig:CPAs2}a, b, and c, respectively) all erupt from PIL 2.  The three CMEs begin at the same longitude, but CME 14D has an initial latitude just under 2\mydeg south of the other two CMEs.  The tilt of CME 14D is 10\mydeg larger (more inclined) than the other two CMEs.  All three CMEs deflect northward and rotate counterclockwise.  CMEs 14C and 16 deflect eastward, and CME 14D initially exhibits a small longitudinal oscillation, similar to CME 14A, before deflecting eastward.  The deflection of CME 14D rotation greatly exceeds that of either CME 14C or 16, which results from the slight difference in its initial position and orientation.

Since these CMEs have such similar initial positions we expect that the variations in their behavior result from either difference in the CME mass or their expansion and radial propagation.  CMEs 14C and 16 both are very light CMEs with masses of only 3$\times$10$^{14}$ g, whereas CME 14D is a slightly heavier 6$\times$10$^{14}$ g.  CME 14C and 16 have the same position, orientation, and mass, yet 14C has a noticeably larger deflection and rotation, which must result from differences in their expansion and radial propagation.  CME 14D's deflection is smaller than that of the other two cases, which can be explained by its larger mass.

\begin{figure}[!hbtp]
\includegraphics[width=4.5in, angle=0]{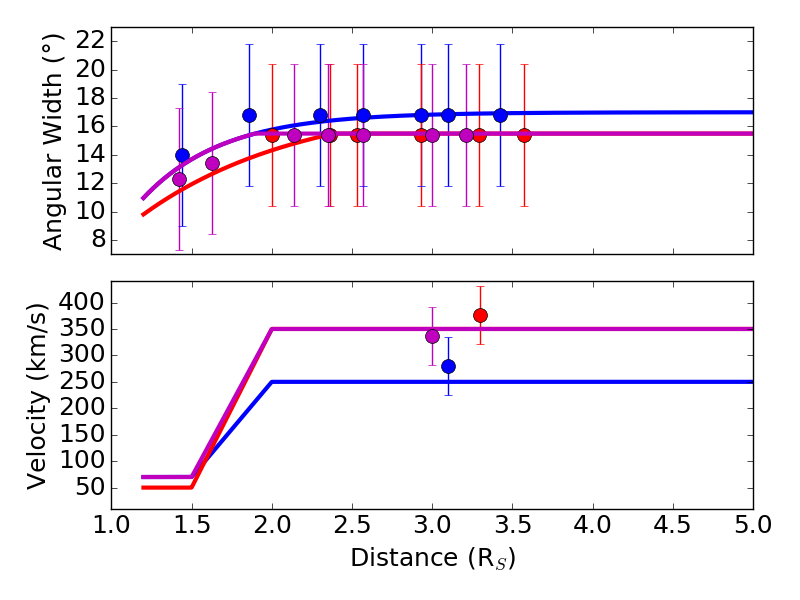}
\caption{Comparison of the expansion (top) and radial propagation (bottom) models used for CMEs 14C (blue), 14D (red), and 16 (purple).  The symbols indicate the angular width reconstructed with the GCS model or the average final radial velocity determined from the farthest three reconstructed positions.}\label{fig:EP}
\end{figure}

Figure \ref{fig:EP} a and b show the expansion and propagation models for CME 14C (blue), 14D (red), and 16 (purple).  These simple empirical functions relating the angular width or radial speed to the radial distance are technically inputs for the ForeCAT model.  The CME angular width and speed are more easily measurable beyond a few solar radii, but precise values are difficult to determine in the low corona.  We use the observed values from the final reconstructed points (typically near 3 to 4 \rsunNS) to constrain the empirical models as much as possible.  Typically we can only constrain the low coronal evolution by determining which values produce ForeCAT results that match the observations.  Figure \ref{fig:EP}a shows the observed angular width for each reconstructed distance with error bars of 5\mydegNS.  Figure \ref{fig:EP}b shows the average final velocity determined from the last three reconstructed points with error bars of 55 \kmsNS, which correspond to an uncertainty in the distance of 0.05 \rsun and a time cadence of 10 minutes.  We do not include the velocity for each position as using a simple finite difference derivative of the radial distance with respect to time yields values that vary by more than physically realistic.  

CMEs 14C and 16 initially have the same angular width and radial velocity.  At 1.5 \rsun both CMEs begin accelerating from their initial radial speeds.  CME 16 has a larger acceleration and reaches a higher final speed.  Just before 2 \rsun CME 16 reaches its maximum angular width and stops overexpanding.  CME 14C continues to slowly increase its angular width, reaching a final angular width slightly larger than that of CME 16.  Both of these effects will cause an increase in the deflection of CME 14C relative to CME 16.  The faster CME 16 spends less time in the region of strong magnetic forces and has a lower density due to its smaller angular width, so it will gain less angular momentum than CME 14C.  Additionally, if two CMEs have the same angular momentum, the one with the slower radial velocity will have more time to deflect before reaching a given radial distance, which will also increase the deflection of CME 14C relative to CME 16.  Figure \ref{fig:CPAs2} shows that CMEs 14C and 16 behave similarly below 1.5 \rsunNS, but their paths diverge when their propagation and expansion differ beyond this distance. 

CME 14D initially has a smaller width than the other two CMEs and a slower initial velocity.  While we would expect an increase in deflection due to the slower initial velocity, the larger CME mass and smaller angular width yield a greater CME density allowing the CME to resist the magnetic deflection forces.  Despite the the smaller deflection, CME 14D undergoes a significant rotation.  We expect that the differences in the expansion and propagation should uniformly affect the deflection and rotation, as for CMEs 14C and 16.  Since the deflection of CME 14D decreases and the rotation increases relative to the other CMEs we expect that this results primarily from the CME position and orientation.  The net force on CME 14D may be more balanced, or the decrease in deflection may result from the increased CME density, but the increase in rotation corresponds to an increase in the torque on the CME.  In \cite{Kay16Obs} we find that the rotation tends to be more sensitive to the small scale, local magnetic gradients, and therefore the initial location and orientation.

\section{Discussion}\label{Disc}

\begin{figure}[!hbtp]
\includegraphics[width=4.5in, angle=0]{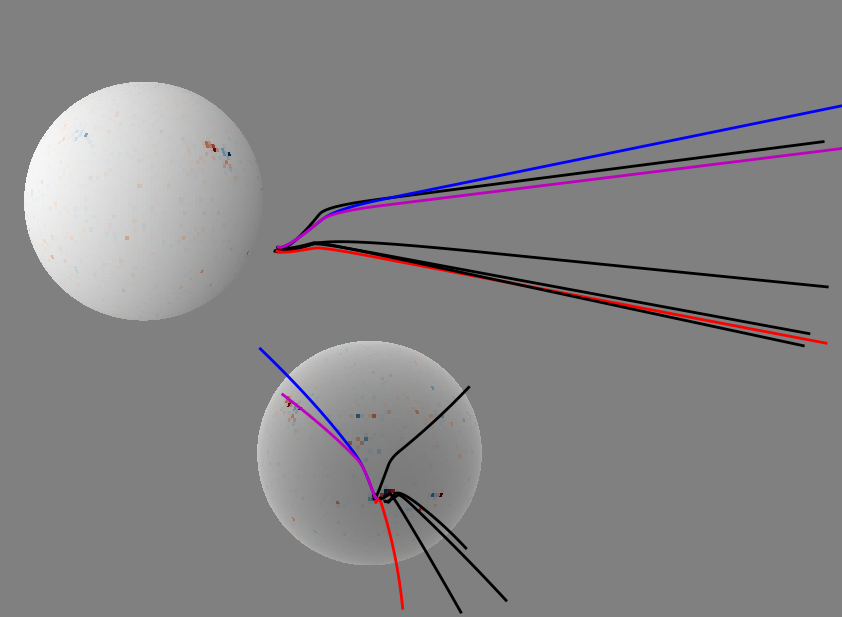}
\caption{A side and front view comparing the trajectories of the seven CMEs out to 6 \rsunNS.  We use the same colors for three CMEs as in Figure \ref{fig:EP}. The surface of the Sun is colored according to the radial magnetic field strength from HMI. A movie version is available in the online supplementary material.}\label{fig:3D}
\end{figure}

To better compare the relative behavior of these CMEs, in Figure \ref{fig:3D} we show the trajectories of all seven cases in three-dimensional space.  Figure \ref{fig:3D} shows one side and one front view, the sphere represents the Sun.  The HMI magnetic field strength has been mapped onto the surface of the Sun.  The online supplementary material includes a movie (AR11158.avi) showing the deflection and rotation of each CME (represented by a blue surface) and comparing their trajectories from different angles.  For each CME the blue surface represents the front of the CME and the black line traces the position of the CME nose.  When we compare the trajectories, three of the CMEs trajectories are colored the same as in Figure \ref{fig:EP}.  The overlap between the positions of CME 14 and CME 16 (blue and purple) close to the Sun can easily be seen in the front view.

At 10 \rsun the latitudes of the CMEs vary between -11.7\mydeg and 7.8\mydeg, and the longitudes vary between 20.5\mydeg and 39.2\mydegNS.  Figure \ref{fig:3D} illustrates that even though all seven CMEs erupted from the same AR they can take very different trajectories, and Figures \ref{fig:CPAs1} and \ref{fig:CPAs2} show that this is evident in both the observations and the simulations.  The observations and simulations also show a wide range in CME rotation with both clockwise and counterclockwise rotations occurring.  All the CMEs from PIL 1 rotate clockwise, and all but one of the four CMEs from PIL 2 rotate counterclockwise. \cite{Tzi13} show that during its evolution AR 11158 builds up large amounts of right-handed helicity, suggesting that the erupting flux ropes would likely all be right-handed.  In this case, only a clockwise rotation should occur due to the internal CME torques.  Two of the simulated counterclockwise rotations of order 5\mydeg and these corresponding observations are consistent with no rotation, however CME 14D shows 30\mydeg of counterclockwise rotation in the observations and the ForeCAT simulation shows that this motion can be reproduced by the external torques.

If space weather forecasters hope to predict CME impacts at Earth it is essential to account for such varied deflections that can occur from a single AR.  Additionally, HMI daily magnetograms show a significant evolution of AR 11158 during the eruptions, however, we were able to reproduce the behavior of all seven CMEs using a single synoptic magnetogram.  We suggest that for these deflections the large scale gradients, which vary less on a day-to-day timescale, must have a stronger influence than the smaller, more rapidly varying magnetic gradients.

Another important aspect for predicting CME impacts is the location of the source region relative to the Earth.  AR 11158 is relatively Earth-directed during 13\,-\,16 February 2011, but due to the roughly 27 day rotation period of the Sun, the active region moves almost 28\mydeg in this time span.  CME 13 erupts from a Stonyhurst longitude of -3.8\mydeg (measured with respect to the Earth longitude), whereas CME 16 erupts from a Stonyhurst longitude of 29.9\mydegNS, despite their Carrington longitudes differing by only 4.1\mydegNS.  We find that the final Stonyhurst longitudes of the CMEs vary between -8.4\mydeg and 20.7\mydegNS.

\cite{Kay16FIDO} suggest that two parameters, the normalized angular distance between the CME nose and the Earth and the relative orientation of the CME tilt and the position angle of the Earth with respect to the CME nose, can be used to predict whether a CME will impact the Earth or not.  CME 15 is the only CME from this article has a counterpart that appears in published lists of interplanetary coronal mass ejections (ICMEs), such as \cite{Ric10} list\footnote{updated list at \url{http://www.srl.caltech.edu/ACE/ASC/DATA/level3/icmetable2.htm}}, despite AR 11158 being relatively Earth-directed the entire time. CME arrived at Earth on 18 February 2011 at 01:30 UT, significantly delayed from the predicted arrival time due to its passage through several smaller CMEs \citealt{Gop13}).  Using the criteria in \cite{Kay16FIDO}, we would only expect CMEs 13, 14A, and 15 to impact the Earth.  CMEs 13 and 14A, however, were very small and slow, so the rapid CME 15 may have reached them and severely disrupted them or amassed them in its swept up solar wind.

\cite{Olm12} also study the 15 February 2011 CME, specifically the interaction between a coronal hole and an EUV wave associated with the CME.  The wave propagating to the south reflects off a coronal hole to the south.  We suggest that the wave behavior may be explained by the motion of the CME edge, which results from a balance of the deflection, rotation, and expansion.  Our simulated CME shows a change in the direction of the southern edge of the CME as it initially deflects northward faster than the bottom edge expands southward, then the motion reverses as the deflection slows and the expansion takes over.  This is not the same motion as in \cite{Olm12}, but our CME is quite simplified and we only consider the net effect of the forces upon a rigid CME torus.  In a more complex treatment, where the individiual sections of the CME could evolve according to the local magnetic forces, the behavior of the CME edges could mirror the EUV wave.

\section{Conclusion}
AR 11158 was one of the most active ARs of Solar Cycle 24, producing 21 flares, including the first X-class flare of the cycle, and 11 CMEs over a span of four days.  Here we use STEREO coronagraph images to determine the trajectory of seven of the CMEs.  These trajectories, combined with our knowledge of the initial location of each CME, show that many of these CME underwent significant deflections and rotations.  Often the reconstructed CME trajectory shows little deflection or rotation beyond 2 \rsunNS, and so that we can only infer that deflection or rotation has occurred based on the difference from the initial CME location and orientation.  We simulate all seven CMEs using ForeCAT, which confirms the expected deflection and rotation between the initial CME position and the closest reconstructed position.  For all seven cases, the ForeCAT results match the reconstructed deflection and rotation within the uncertainty of the reconstruction technique.  

The CMEs can be separated into eruptions from two different PILs.  All CMEs show a northward deflection, with the magnitude ranging between 9\mydeg and 27\mydegNS, which tends to scale with CME mass.  Most of the CMEs exhibit small longitudinal deflections (less than approximately 5\mydegNS), but two of the CMEs (both from the more eastward PIL) show significant eastward deflections of order 10\mydegNS.  The CME rotations vary in magnitude between 5\mydeg and 50\mydeg and occur in both the clockwise and counterclockwise direction from both PILs.

Despite all originating from a single AR, we find very different behavior for the seven CMEs.  The uniform northward motion corresponds to deflection toward the HCS.  We expect that the large latitudinal deflections result from this coherent force integrated over the CME, whereas the smaller longitudinal motion results from smaller scale magnetic gradients related to the structure of the AR.  For most of these CMEs, the varied longitudinal forces tend to balance out when averaged over the CME.  For three CMEs with very similiar initial positions and orientations, we find that the differences in their deflections and rotations can be explained by differences in their expansion, radial propagation, and mass.  Finally, we suggest that the deflection and rotation can explain why only one of the seven CMEs had clear \textit{in situ} signatures near Earth, despite the AR facing Earth for all of the eruptions. 

\acknowledgements
C.K.'s research was supported by an appointment to the NASA Postdoctoral Program at NASA GSFC, administered by the Universities Space Research Association under contract with NASA. H. X. was partially supported by NASA grant NNX15AB70G. Disclosure of Potential Conflicts of Interest The authors declare that they have no conflicts of interest.



\begin{thebibliography}{41}
\ifx\bisbn     \undefined \def\bisbn  #1{ISBN #1}\fi
\ifx\binits    \undefined \def\binits#1{#1}\fi
\ifx\bauthor   \undefined \def\bauthor#1{#1}\fi
\ifx\batitle   \undefined \def\batitle#1{#1}\fi
\ifx\bjtitle   \undefined \def\bjtitle#1{\textit{#1}}\fi
\ifx\bvolume   \undefined \def\bvolume#1{\textbf{#1}}\fi
\ifx\byear     \undefined \def\byear#1{#1}\fi
\ifx\bissue    \undefined \def\bissue#1{#1}\fi
\ifx\bfpage    \undefined \def\bfpage#1{#1}\fi
\ifx\blpage    \undefined \def\blpage #1{#1}\fi
\ifx\burl      \undefined \def\burl#1{\textsf{#1}}\fi
\ifx\href      \undefined \def\href#1#2{\textsf{#2}}\fi
\ifx\betal     \undefined \def\betal{\textit{et al.}}\fi
\ifx\bctitle   \undefined \def\bctitle#1{#1}\fi
\ifx\beditor   \undefined \def\beditor#1{#1}\fi
\ifx\bbtitle   \undefined \def\bbtitle#1{\textit{#1}}\fi
\ifx\bedition  \undefined \def\bedition#1{#1}\fi
\ifx\bseriesno \undefined \def\bseriesno#1{\textbf{#1}}\fi
\ifx\blocation \undefined \def\blocation#1{#1}\fi
\ifx\bsertitle \undefined \def\bsertitle#1{\textit{#1}}\fi
\ifx\bsnm      \undefined \def\bsnm#1{#1}\fi
\ifx\bsuffix   \undefined \def\bsuffix#1{#1}\fi
\ifx\bparticle \undefined \def\bparticle#1{#1}\fi
\ifx\barticle  \undefined \def\barticle#1{}\fi
\ifx\binstitute  \undefined \def\binstitute#1{#1}\fi
\ifx\bpublisher  \undefined \def\bpublisher#1{#1}\fi
\ifx\doiurl    \undefined
  \def\doiurl#1{\href{http://dx.doi.org/#1}{\textsf{DOI}}}\fi
\ifx\arxivurl  \undefined
  \def\arxivurl#1{\href{http://arxiv.org/abs/#1}{\textsf{arXiv}}}\fi
\ifx\adsurl    \undefined
  \def\adsurl#1{\href{http://adsabs.harvard.edu/abs/#1}{\textsf{ADS}}}\fi
\ifx\botherref \undefined \def\botherref#1{}\fi
\ifx\url       \undefined \def\url#1{\textsf{#1}}\fi
\ifx\bchapter  \undefined \def\bchapter#1{}\fi
\ifx\bbook     \undefined \def\bbook#1{}\fi
\ifx\bcomment  \undefined \def\bcomment#1{#1}\fi
\ifx\oauthor   \undefined \def\oauthor#1{#1}\fi
\ifx\citeauthoryear \undefined\def \citeauthoryear#1{#1}\fi
\ifx\endbibitem\undefined \def\endbibitem{}\fi
\ifx\bconflocation  \undefined \def\bconflocation#1{#1} \fi

\bibitem[\protect\citeauthoryear{{Altschuler} and {Newkirk}}{1969}]{Alt69}
\begin{barticle}
\bauthor{\bsnm{{Altschuler}}, \binits{M.D.}},
\bauthor{\bsnm{{Newkirk}}, \binits{G.}}:
\byear{1969},
\batitle{{Magnetic Fields and the Structure of the Solar Corona. I: Methods of
  Calculating Coronal Fields}}.
\bjtitle{\solphys}
\bvolume{9},
\bfpage{131}.
\doiurl{10.1007/BF00145734}.
\adsurl{1969SoPh....9..131A}.
\end{barticle}
\endbibitem

\bibitem[\protect\citeauthoryear{{Cremades} and {Bothmer}}{2004}]{Cre04}
\begin{barticle}
\bauthor{\bsnm{{Cremades}}, \binits{H.}},
\bauthor{\bsnm{{Bothmer}}, \binits{V.}}:
\byear{2004},
\batitle{{On the three-dimensional configuration of coronal mass ejections}}.
\bjtitle{\aap}
\bvolume{422},
\bfpage{307}.
\doiurl{10.1051/0004-6361:20035776}.
\adsurl{2004A\%26A...422..307C}.
\end{barticle}
\endbibitem

\bibitem[\protect\citeauthoryear{{Fan} and {Gibson}}{2004}]{Fan04}
\begin{barticle}
\bauthor{\bsnm{{Fan}}, \binits{Y.}},
\bauthor{\bsnm{{Gibson}}, \binits{S.E.}}:
\byear{2004},
\batitle{{Numerical Simulations of Three-dimensional Coronal Magnetic Fields
  Resulting from the Emergence of Twisted Magnetic Flux Tubes}}.
\bjtitle{\apj}
\bvolume{609},
\bfpage{1123}.
\doiurl{10.1086/421238}.
\adsurl{2004ApJ...609.1123F}.
\end{barticle}
\endbibitem

\bibitem[\protect\citeauthoryear{{Filippov}, {Gopalswamy}, and
  {Lozhechkin}}{2001}]{Fil01}
\begin{barticle}
\bauthor{\bsnm{{Filippov}}, \binits{B.P.}},
\bauthor{\bsnm{{Gopalswamy}}, \binits{N.}},
\bauthor{\bsnm{{Lozhechkin}}, \binits{A.V.}}:
\byear{2001},
\batitle{{Non-radial motion of eruptive filaments}}.
\bjtitle{\solphys}
\bvolume{203},
\bfpage{119}.
\doiurl{10.1023/A:1012754329767}.
\adsurl{2001SoPh..203..119F}.
\end{barticle}
\endbibitem

\bibitem[\protect\citeauthoryear{{Gopalswamy} \textit{et~al.}}{2009}]{Gop09}
\begin{barticle}
\bauthor{\bsnm{{Gopalswamy}}, \binits{N.}},
\bauthor{\bsnm{{M{\"a}kel{\"a}}}, \binits{P.}},
\bauthor{\bsnm{{Xie}}, \binits{H.}},
\bauthor{\bsnm{{Akiyama}}, \binits{S.}},
\bauthor{\bsnm{{Yashiro}}, \binits{S.}}:
\byear{2009},
\batitle{{CME interactions with coronal holes and their interplanetary
  consequences}}.
\bjtitle{J. Geophys. Res. (Space Physics)}
\bvolume{114},
\bfpage{A00A22}.
\doiurl{10.1029/2008JA013686}.
\adsurl{2009JGRA..114.0A22G}.
\end{barticle}
\endbibitem

\bibitem[\protect\citeauthoryear{{Gopalswamy} \textit{et~al.}}{2013}]{Gop13}
\begin{barticle}
\bauthor{\bsnm{{Gopalswamy}}, \binits{N.}},
\bauthor{\bsnm{{M{\"a}kel{\"a}}}, \binits{P.}},
\bauthor{\bsnm{{Xie}}, \binits{H.}},
\bauthor{\bsnm{{Yashiro}}, \binits{S.}}:
\byear{2013},
\batitle{{Testing the empirical shock arrival model using quadrature
  observations}}.
\bjtitle{Space Weather}
\bvolume{11},
\bfpage{661}.
\doiurl{10.1002/2013SW000945}.
\adsurl{2013SpWea..11..661G}.
\end{barticle}
\endbibitem

\bibitem[\protect\citeauthoryear{{Gosling} \textit{et~al.}}{2005}]{Gos05}
\begin{barticle}
\bauthor{\bsnm{{Gosling}}, \binits{J.T.}},
\bauthor{\bsnm{{Skoug}}, \binits{R.M.}},
\bauthor{\bsnm{{McComas}}, \binits{D.J.}},
\bauthor{\bsnm{{Smith}}, \binits{C.W.}}:
\byear{2005},
\batitle{{Magnetic disconnection from the Sun: Observations of a reconnection
  exhaust in the solar wind at the heliospheric current sheet}}.
\bjtitle{\grl}
\bvolume{32},
\bfpage{L05105}.
\doiurl{10.1029/2005GL022406}.
\adsurl{2005GeoRL..32.5105G}.
\end{barticle}
\endbibitem

\bibitem[\protect\citeauthoryear{{Gui} \textit{et~al.}}{2011}]{Gui11}
\begin{barticle}
\bauthor{\bsnm{{Gui}}, \binits{B.}},
\bauthor{\bsnm{{Shen}}, \binits{C.}},
\bauthor{\bsnm{{Wang}}, \binits{Y.}},
\bauthor{\bsnm{{Ye}}, \binits{P.}},
\bauthor{\bsnm{{Liu}}, \binits{J.}},
\bauthor{\bsnm{{Wang}}, \binits{S.}},
\bauthor{\bsnm{{Zhao}}, \binits{X.}}:
\byear{2011},
\batitle{{Quantitative Analysis of CME Deflections in the Corona}}.
\bjtitle{\solphys}
\bvolume{271},
\bfpage{111}.
\doiurl{10.1007/s11207-011-9791-9}.
\adsurl{2011SoPh..271..111G}.
\end{barticle}
\endbibitem

\bibitem[\protect\citeauthoryear{{Hildner}}{1977}]{Hil77}
\begin{bchapter}
\bauthor{\bsnm{{Hildner}}, \binits{E.}}:
\byear{1977},
\bctitle{{Mass ejections from the solar corona into interplanetary space}}.
In: \beditor{\bsnm{{Shea}}, \binits{M.A.}},
\beditor{\bsnm{{Smart}}, \binits{D.F.}},
\beditor{\bsnm{{Wu}}, \binits{S.T.}} (eds.)
\bbtitle{Study of Travelling Interplanetary Phenomena},
\bsertitle{Astrophys. and Space Sci. Lib.}
\bseriesno{71},
\bfpage{3}.
\adsurl{1977ASSL...71....3H}.
\end{bchapter}
\endbibitem

\bibitem[\protect\citeauthoryear{{Isavnin}, {Vourlidas}, and
  {Kilpua}}{2013}]{Isa13}
\begin{barticle}
\bauthor{\bsnm{{Isavnin}}, \binits{A.}},
\bauthor{\bsnm{{Vourlidas}}, \binits{A.}},
\bauthor{\bsnm{{Kilpua}}, \binits{E.K.J.}}:
\byear{2013},
\batitle{{Three-Dimensional Evolution of Erupted Flux Ropes from the Sun (2 -
  20 R $_S$) to 1 AU}}.
\bjtitle{\solphys}
\bvolume{284},
\bfpage{203}.
\doiurl{10.1007/s11207-012-0214-3}.
\adsurl{2013SoPh..284..203I}.
\end{barticle}
\endbibitem

\bibitem[\protect\citeauthoryear{{Kay} and {Opher}}{2015}]{Kay15AM}
\begin{barticle}
\bauthor{\bsnm{{Kay}}, \binits{C.}},
\bauthor{\bsnm{{Opher}}, \binits{M.}}:
\byear{2015},
\batitle{{The Heliocentric Distance where the Deflections and Rotations of
  Solar Coronal Mass Ejections Occur}}.
\bjtitle{\apjl}
\bvolume{811},
\bfpage{L36}.
\doiurl{10.1088/2041-8205/811/2/L36}.
\adsurl{2015ApJ...811L..36K}.
\end{barticle}
\endbibitem

\bibitem[\protect\citeauthoryear{{Kay}, {dos Santos}, and
  {Opher}}{2015}]{Kay15L}
\begin{barticle}
\bauthor{\bsnm{{Kay}}, \binits{C.}},
\bauthor{\bsnm{{dos Santos}}, \binits{L.F.G.}},
\bauthor{\bsnm{{Opher}}, \binits{M.}}:
\byear{2015},
\batitle{{Constraining the Masses and the Non-radial Drag Coefficient of a
  Solar Coronal Mass Ejection}}.
\bjtitle{\apjl}
\bvolume{801},
\bfpage{L21}.
\doiurl{10.1088/2041-8205/801/2/L21}.
\adsurl{2015ApJ...801L..21K}.
\end{barticle}
\endbibitem

\bibitem[\protect\citeauthoryear{{Kay}, {Opher}, and {Evans}}{2013}]{Kay13}
\begin{barticle}
\bauthor{\bsnm{{Kay}}, \binits{C.}},
\bauthor{\bsnm{{Opher}}, \binits{M.}},
\bauthor{\bsnm{{Evans}}, \binits{R.M.}}:
\byear{2013},
\batitle{{Forecasting a Coronal Mass Ejection's Altered Trajectory: ForeCAT}}.
\bjtitle{\apj}
\bvolume{775},
\bfpage{5}.
\doiurl{10.1088/0004-637X/775/1/5}.
\adsurl{2013ApJ...775....5K}.
\end{barticle}
\endbibitem

\bibitem[\protect\citeauthoryear{{Kay}, {Opher}, and {Evans}}{2015}]{Kay15}
\begin{barticle}
\bauthor{\bsnm{{Kay}}, \binits{C.}},
\bauthor{\bsnm{{Opher}}, \binits{M.}},
\bauthor{\bsnm{{Evans}}, \binits{R.M.}}:
\byear{2015},
\batitle{{Global Trends of CME Deflections Based on CME and Solar Parameters}}.
\bjtitle{\apj}
\bvolume{805},
\bfpage{168}.
\doiurl{10.1088/0004-637X/805/2/168}.
\adsurl{2015ApJ...805..168K}.
\end{barticle}
\endbibitem

\bibitem[\protect\citeauthoryear{{Kay} \textit{et~al.}}{2016a}]{Kay16FIDO}
\begin{botherref}
\oauthor{\bsnm{{Kay}}, \binits{C.}},
\oauthor{\bsnm{{Gopalswamy}}, \binits{N.}},
\oauthor{\bsnm{{Reinard}}, \binits{A.}},
\oauthor{\bsnm{{Opher}}, \binits{M.}}:
2016a,
{Determining ICME Magnetic Field Orientations with the ForeCAT In situ Data
  Observer}.
\textit{\apj}.
\end{botherref}
\endbibitem

\bibitem[\protect\citeauthoryear{{Kay} \textit{et~al.}}{2016b}]{Kay16Obs}
\begin{barticle}
\bauthor{\bsnm{{Kay}}, \binits{C.}},
\bauthor{\bsnm{{Opher}}, \binits{M.}},
\bauthor{\bsnm{{Colaninno}}, \binits{R.C.}},
\bauthor{\bsnm{{Vourlidas}}, \binits{A.}}:
\byear{2016}b,
\batitle{{Using ForeCAT Deflections and Rotations to Constrain the Early
  Evolution of CMEs}}.
\bjtitle{\apj}
\bvolume{827},
\bfpage{70}.
\doiurl{10.3847/0004-637X/827/1/70}.
\adsurl{2016ApJ...827...70K}.
\end{barticle}
\endbibitem

\bibitem[\protect\citeauthoryear{{Kilpua} \textit{et~al.}}{2009}]{Kil09}
\begin{barticle}
\bauthor{\bsnm{{Kilpua}}, \binits{E.K.J.}},
\bauthor{\bsnm{{Pomoell}}, \binits{J.}},
\bauthor{\bsnm{{Vourlidas}}, \binits{A.}},
\bauthor{\bsnm{{Vainio}}, \binits{R.}},
\bauthor{\bsnm{{Luhmann}}, \binits{J.}},
\bauthor{\bsnm{{Li}}, \binits{Y.}},
\bauthor{\bsnm{{Schroeder}}, \binits{P.}},
\bauthor{\bsnm{{Galvin}}, \binits{A.B.}},
\bauthor{\bsnm{{Simunac}}, \binits{K.}}:
\byear{2009},
\batitle{{STEREO observations of interplanetary coronal mass ejections and
  prominence deflection during solar minimum period}}.
\bjtitle{Ann. Geophys.}
\bvolume{27},
\bfpage{4491}.
\doiurl{10.5194/angeo-27-4491-2009}.
\adsurl{2009AnGeo..27.4491K}.
\end{barticle}
\endbibitem

\bibitem[\protect\citeauthoryear{{Lemen} \textit{et~al.}}{2012}]{AIA}
\begin{barticle}
\bauthor{\bsnm{{Lemen}}, \binits{J.R.}},
\bauthor{\bsnm{{Title}}, \binits{A.M.}},
\bauthor{\bsnm{{Akin}}, \binits{D.J.}},
\bauthor{\bsnm{{Boerner}}, \binits{P.F.}},
\bauthor{\bsnm{{Chou}}, \binits{C.}},
\bauthor{\bsnm{{Drake}}, \binits{J.F.}},
\bauthor{\bsnm{\textit{et al.}}}:
\byear{2012},
\batitle{{The Atmospheric Imaging Assembly (AIA) on the Solar Dynamics
  Observatory (SDO)}}.
\bjtitle{\solphys}
\bvolume{275},
\bfpage{17}.
\doiurl{10.1007/s11207-011-9776-8}.
\adsurl{2012SoPh..275...17L}.
\end{barticle}
\endbibitem

\bibitem[\protect\citeauthoryear{{Liewer} \textit{et~al.}}{2015}]{Lie15}
\begin{barticle}
\bauthor{\bsnm{{Liewer}}, \binits{P.}},
\bauthor{\bsnm{{Panasenco}}, \binits{O.}},
\bauthor{\bsnm{{Vourlidas}}, \binits{A.}},
\bauthor{\bsnm{{Colaninno}}, \binits{R.}}:
\byear{2015},
\batitle{{Observations and Analysis of the Non-Radial Propagation of Coronal
  Mass Ejections Near the Sun}}.
\bjtitle{\solphys}
\bvolume{290},
\bfpage{3343}.
\doiurl{10.1007/s11207-015-0794-9}.
\adsurl{2015SoPh..290.3343L}.
\end{barticle}
\endbibitem

\bibitem[\protect\citeauthoryear{{Lynch} \textit{et~al.}}{2009}]{Lyn09}
\begin{barticle}
\bauthor{\bsnm{{Lynch}}, \binits{B.J.}},
\bauthor{\bsnm{{Antiochos}}, \binits{S.K.}},
\bauthor{\bsnm{{Li}}, \binits{Y.}},
\bauthor{\bsnm{{Luhmann}}, \binits{J.G.}},
\bauthor{\bsnm{{DeVore}}, \binits{C.R.}}:
\byear{2009},
\batitle{{Rotation of Coronal Mass Ejections during Eruption}}.
\bjtitle{\apj}
\bvolume{697},
\bfpage{1918}.
\doiurl{10.1088/0004-637X/697/2/1918}.
\adsurl{2009ApJ...697.1918L}.
\end{barticle}
\endbibitem

\bibitem[\protect\citeauthoryear{{MacQueen}, {Hundhausen}, and
  {Conover}}{1986}]{Mac86}
\begin{barticle}
\bauthor{\bsnm{{MacQueen}}, \binits{R.M.}},
\bauthor{\bsnm{{Hundhausen}}, \binits{A.J.}},
\bauthor{\bsnm{{Conover}}, \binits{C.W.}}:
\byear{1986},
\batitle{{The propagation of coronal mass ejection transients}}.
\bjtitle{\jgr}
\bvolume{91},
\bfpage{31}.
\doiurl{10.1029/JA091iA01p00031}.
\adsurl{1986JGR....91...31M}.
\end{barticle}
\endbibitem

\bibitem[\protect\citeauthoryear{{Mays} \textit{et~al.}}{2015}]{May15AT}
\begin{barticle}
\bauthor{\bsnm{{Mays}}, \binits{M.L.}},
\bauthor{\bsnm{{Taktakishvili}}, \binits{A.}},
\bauthor{\bsnm{{Pulkkinen}}, \binits{A.}},
\bauthor{\bsnm{{MacNeice}}, \binits{P.J.}},
\bauthor{\bsnm{{Rast{\"a}tter}}, \binits{L.}},
\bauthor{\bsnm{{Odstrcil}}, \binits{D.}},
\bauthor{\bsnm{{Jian}}, \binits{L.K.}},
\bauthor{\bsnm{{Richardson}}, \binits{I.G.}},
\bauthor{\bsnm{{LaSota}}, \binits{J.A.}},
\bauthor{\bsnm{{Zheng}}, \binits{Y.}},
\bauthor{\bsnm{{Kuznetsova}}, \binits{M.M.}}:
\byear{2015},
\batitle{{Ensemble Modeling of CMEs Using the WSA-ENLIL+Cone Model}}.
\bjtitle{\solphys}
\bvolume{290},
\bfpage{1775}.
\doiurl{10.1007/s11207-015-0692-1}.
\adsurl{2015SoPh..290.1775M}.
\end{barticle}
\endbibitem

\bibitem[\protect\citeauthoryear{{Nieves-Chinchilla}
  \textit{et~al.}}{2012}]{Nie12}
\begin{barticle}
\bauthor{\bsnm{{Nieves-Chinchilla}}, \binits{T.}},
\bauthor{\bsnm{{Colaninno}}, \binits{R.}},
\bauthor{\bsnm{{Vourlidas}}, \binits{A.}},
\bauthor{\bsnm{{Szabo}}, \binits{A.}},
\bauthor{\bsnm{{Lepping}}, \binits{R.P.}},
\bauthor{\bsnm{{Boardsen}}, \binits{S.A.}},
\bauthor{\bsnm{{Anderson}}, \binits{B.J.}},
\bauthor{\bsnm{{Korth}}, \binits{H.}}:
\byear{2012},
\batitle{{Remote and in situ observations of an unusual Earth-directed coronal
  mass ejection from multiple viewpoints}}.
\bjtitle{J. of Geophys. Res. (Space Physics)}
\bvolume{117},
\bfpage{6106}.
\doiurl{10.1029/2011JA017243}.
\adsurl{2012JGRA..117.6106N}.
\end{barticle}
\endbibitem

\bibitem[\protect\citeauthoryear{{Nieves-Chinchilla}
  \textit{et~al.}}{2013}]{Nie13}
\begin{barticle}
\bauthor{\bsnm{{Nieves-Chinchilla}}, \binits{T.}},
\bauthor{\bsnm{{Vourlidas}}, \binits{A.}},
\bauthor{\bsnm{{Stenborg}}, \binits{G.}},
\bauthor{\bsnm{{Savani}}, \binits{N.P.}},
\bauthor{\bsnm{{Koval}}, \binits{A.}},
\bauthor{\bsnm{{Szabo}}, \binits{A.}},
\bauthor{\bsnm{{Jian}}, \binits{L.K.}}:
\byear{2013},
\batitle{{Inner Heliospheric Evolution of a ''Stealth'' CME Derived from
  Multi-view Imaging and Multipoint in Situ observations. I. Propagation to 1
  AU}}.
\bjtitle{\apj}
\bvolume{779},
\bfpage{55}.
\doiurl{10.1088/0004-637X/779/1/55}.
\adsurl{2013ApJ...779...55N}.
\end{barticle}
\endbibitem

\bibitem[\protect\citeauthoryear{{Olmedo} \textit{et~al.}}{2012}]{Olm12}
\begin{barticle}
\bauthor{\bsnm{{Olmedo}}, \binits{O.}},
\bauthor{\bsnm{{Vourlidas}}, \binits{A.}},
\bauthor{\bsnm{{Zhang}}, \binits{J.}},
\bauthor{\bsnm{{Cheng}}, \binits{X.}}:
\byear{2012},
\batitle{{Secondary Waves and/or the ''Reflection'' from and ''Transmission''
  through a Coronal Hole of an Extreme Ultraviolet Wave Associated with the
  2011 February 15 X2.2 Flare Observed with SDO/AIA and STEREO/EUVI}}.
\bjtitle{\apj}
\bvolume{756},
\bfpage{143}.
\doiurl{10.1088/0004-637X/756/2/143}.
\adsurl{2012ApJ...756..143O}.
\end{barticle}
\endbibitem

\bibitem[\protect\citeauthoryear{{Panasenco} \textit{et~al.}}{2011}]{Pan11}
\begin{barticle}
\bauthor{\bsnm{{Panasenco}}, \binits{O.}},
\bauthor{\bsnm{{Martin}}, \binits{S.}},
\bauthor{\bsnm{{Joshi}}, \binits{A.D.}},
\bauthor{\bsnm{{Srivastava}}, \binits{N.}}:
\byear{2011},
\batitle{{Rolling motion in erupting prominences observed by STEREO}}.
\bjtitle{J. of Atmos. and Solar-Terr. Phys.}
\bvolume{73},
\bfpage{1129}.
\doiurl{10.1016/j.jastp.2010.09.010}.
\adsurl{2011JASTP..73.1129P}.
\end{barticle}
\endbibitem

\bibitem[\protect\citeauthoryear{{Panasenco} \textit{et~al.}}{2013}]{Pan13}
\begin{barticle}
\bauthor{\bsnm{{Panasenco}}, \binits{O.}},
\bauthor{\bsnm{{Martin}}, \binits{S.F.}},
\bauthor{\bsnm{{Velli}}, \binits{M.}},
\bauthor{\bsnm{{Vourlidas}}, \binits{A.}}:
\byear{2013},
\batitle{{Origins of Rolling, Twisting, and Non-radial Propagation of Eruptive
  Solar Events}}.
\bjtitle{\solphys}
\bvolume{287},
\bfpage{391}.
\doiurl{10.1007/s11207-012-0194-3}.
\adsurl{2013SoPh..287..391P}.
\end{barticle}
\endbibitem

\bibitem[\protect\citeauthoryear{{Pesnell}, {Thompson}, and
  {Chamberlin}}{2012}]{SDO}
\begin{barticle}
\bauthor{\bsnm{{Pesnell}}, \binits{W.D.}},
\bauthor{\bsnm{{Thompson}}, \binits{B.J.}},
\bauthor{\bsnm{{Chamberlin}}, \binits{P.C.}}:
\byear{2012},
\batitle{{The Solar Dynamics Observatory (SDO)}}.
\bjtitle{\solphys}
\bvolume{275},
\bfpage{3}.
\doiurl{10.1007/s11207-011-9841-3}.
\adsurl{2012SoPh..275....3P}.
\end{barticle}
\endbibitem

\bibitem[\protect\citeauthoryear{{Richardson} and {Cane}}{2010}]{Ric10}
\begin{barticle}
\bauthor{\bsnm{{Richardson}}, \binits{I.G.}},
\bauthor{\bsnm{{Cane}}, \binits{H.V.}}:
\byear{2010},
\batitle{{Near-Earth Interplanetary Coronal Mass Ejections During Solar Cycle
  23 (1996 - 2009): Catalog and Summary of Properties}}.
\bjtitle{\solphys}
\bvolume{264},
\bfpage{189}.
\doiurl{10.1007/s11207-010-9568-6}.
\adsurl{2010SoPh..264..189R}.
\end{barticle}
\endbibitem

\bibitem[\protect\citeauthoryear{{Savani} \textit{et~al.}}{2010}]{Sav10}
\begin{barticle}
\bauthor{\bsnm{{Savani}}, \binits{N.P.}},
\bauthor{\bsnm{{Owens}}, \binits{M.J.}},
\bauthor{\bsnm{{Rouillard}}, \binits{A.P.}},
\bauthor{\bsnm{{Forsyth}}, \binits{R.J.}},
\bauthor{\bsnm{{Davies}}, \binits{J.A.}}:
\byear{2010},
\batitle{{Observational Evidence of a Coronal Mass Ejection Distortion Directly
  Attributable to a Structured Solar Wind}}.
\bjtitle{\apjl}
\bvolume{714},
\bfpage{L128}.
\doiurl{10.1088/2041-8205/714/1/L128}.
\adsurl{2010ApJ...714L.128S}.
\end{barticle}
\endbibitem

\bibitem[\protect\citeauthoryear{{Schatten}, {Wilcox}, and
  {Ness}}{1969}]{Sch69}
\begin{barticle}
\bauthor{\bsnm{{Schatten}}, \binits{K.H.}},
\bauthor{\bsnm{{Wilcox}}, \binits{J.M.}},
\bauthor{\bsnm{{Ness}}, \binits{N.F.}}:
\byear{1969},
\batitle{{A model of interplanetary and coronal magnetic fields}}.
\bjtitle{\solphys}
\bvolume{6},
\bfpage{442}.
\doiurl{10.1007/BF00146478}.
\adsurl{1969SoPh....6..442S}.
\end{barticle}
\endbibitem

\bibitem[\protect\citeauthoryear{{Schou} \textit{et~al.}}{2012}]{HMI}
\begin{barticle}
\bauthor{\bsnm{{Schou}}, \binits{J.}},
\bauthor{\bsnm{{Scherrer}}, \binits{P.H.}},
\bauthor{\bsnm{{Bush}}, \binits{R.I.}},
\bauthor{\bsnm{{Wachter}}, \binits{R.}},
\bauthor{\bsnm{{Couvidat}}, \binits{S.}},
\bauthor{\bsnm{{Rabello-Soares}}, \binits{M.C.}},
\bauthor{\bsnm{{Bogart}}, \binits{R.S.}},
\bauthor{\bsnm{{Hoeksema}}, \binits{J.T.}},
\bauthor{\bsnm{{Liu}}, \binits{Y.}},
\bauthor{\bsnm{{Duvall}}, \binits{T.L.}},
\bauthor{\bsnm{{Akin}}, \binits{D.J.}},
\bauthor{\bsnm{{Allard}}, \binits{B.A.}},
\bauthor{\bsnm{{Miles}}, \binits{J.W.}},
\bauthor{\bsnm{{Rairden}}, \binits{R.}},
\bauthor{\bsnm{{Shine}}, \binits{R.A.}},
\bauthor{\bsnm{{Tarbell}}, \binits{T.D.}},
\bauthor{\bsnm{{Title}}, \binits{A.M.}},
\bauthor{\bsnm{{Wolfson}}, \binits{C.J.}},
\bauthor{\bsnm{{Elmore}}, \binits{D.F.}},
\bauthor{\bsnm{{Norton}}, \binits{A.A.}},
\bauthor{\bsnm{{Tomczyk}}, \binits{S.}}:
\byear{2012},
\batitle{{Design and Ground Calibration of the Helioseismic and Magnetic Imager
  (HMI) Instrument on the Solar Dynamics Observatory (SDO)}}.
\bjtitle{\solphys}
\bvolume{275},
\bfpage{229}.
\doiurl{10.1007/s11207-011-9842-2}.
\adsurl{2012SoPh..275..229S}.
\end{barticle}
\endbibitem

\bibitem[\protect\citeauthoryear{{Shen} \textit{et~al.}}{2011}]{She11}
\begin{barticle}
\bauthor{\bsnm{{Shen}}, \binits{C.}},
\bauthor{\bsnm{{Wang}}, \binits{Y.}},
\bauthor{\bsnm{{Gui}}, \binits{B.}},
\bauthor{\bsnm{{Ye}}, \binits{P.}},
\bauthor{\bsnm{{Wang}}, \binits{S.}}:
\byear{2011},
\batitle{{Kinematic Evolution of a Slow CME in Corona Viewed by STEREO-B on 8
  October 2007}}.
\bjtitle{\solphys}
\bvolume{269},
\bfpage{389}.
\doiurl{10.1007/s11207-011-9715-8}.
\adsurl{2011SoPh..269..389S}.
\end{barticle}
\endbibitem

\bibitem[\protect\citeauthoryear{{Thernisien}, {Vourlidas}, and
  {Howard}}{2009}]{The09}
\begin{barticle}
\bauthor{\bsnm{{Thernisien}}, \binits{A.}},
\bauthor{\bsnm{{Vourlidas}}, \binits{A.}},
\bauthor{\bsnm{{Howard}}, \binits{R.A.}}:
\byear{2009},
\batitle{{Forward Modeling of Coronal Mass Ejections Using STEREO/SECCHI
  Data}}.
\bjtitle{\solphys}
\bvolume{256},
\bfpage{111}.
\doiurl{10.1007/s11207-009-9346-5}.
\adsurl{2009SoPh..256..111T}.
\end{barticle}
\endbibitem

\bibitem[\protect\citeauthoryear{{Thernisien}, {Howard}, and
  {Vourlidas}}{2006}]{The06}
\begin{barticle}
\bauthor{\bsnm{{Thernisien}}, \binits{A.F.R.}},
\bauthor{\bsnm{{Howard}}, \binits{R.A.}},
\bauthor{\bsnm{{Vourlidas}}, \binits{A.}}:
\byear{2006},
\batitle{{Modeling of Flux Rope Coronal Mass Ejections}}.
\bjtitle{\apj}
\bvolume{652},
\bfpage{763}.
\doiurl{10.1086/508254}.
\adsurl{2006ApJ...652..763T}.
\end{barticle}
\endbibitem

\bibitem[\protect\citeauthoryear{{Thompson}, {Kliem}, and
  {T{\"o}r{\"o}k}}{2012}]{Tho12}
\begin{barticle}
\bauthor{\bsnm{{Thompson}}, \binits{W.T.}},
\bauthor{\bsnm{{Kliem}}, \binits{B.}},
\bauthor{\bsnm{{T{\"o}r{\"o}k}}, \binits{T.}}:
\byear{2012},
\batitle{{3D Reconstruction of a Rotating Erupting Prominence}}.
\bjtitle{\solphys}
\bvolume{276},
\bfpage{241}.
\doiurl{10.1007/s11207-011-9868-5}.
\adsurl{2012SoPh..276..241T}.
\end{barticle}
\endbibitem

\bibitem[\protect\citeauthoryear{{Thompson} \textit{et~al.}}{2003}]{COR1}
\begin{bchapter}
\bauthor{\bsnm{{Thompson}}, \binits{W.T.}},
\bauthor{\bsnm{{Davila}}, \binits{J.M.}},
\bauthor{\bsnm{{Fisher}}, \binits{R.R.}},
\bauthor{\bsnm{{Orwig}}, \binits{L.E.}},
\bauthor{\bsnm{{Mentzell}}, \binits{J.E.}},
\bauthor{\bsnm{{Hetherington}}, \binits{S.E.}},
\bauthor{\bsnm{{Derro}}, \binits{R.J.}},
\bauthor{\bsnm{{Federline}}, \binits{R.E.}},
\bauthor{\bsnm{{Clark}}, \binits{D.C.}},
\bauthor{\bsnm{{Chen}}, \binits{P.T.C.}},
\bauthor{\bsnm{{Tveekrem}}, \binits{J.L.}},
\bauthor{\bsnm{{Martino}}, \binits{A.J.}},
\bauthor{\bsnm{{Novello}}, \binits{J.}},
\bauthor{\bsnm{{Wesenberg}}, \binits{R.P.}},
\bauthor{\bsnm{{StCyr}}, \binits{O.C.}},
\bauthor{\bsnm{{Reginald}}, \binits{N.L.}},
\bauthor{\bsnm{{Howard}}, \binits{R.A.}},
\bauthor{\bsnm{{Mehalick}}, \binits{K.I.}},
\bauthor{\bsnm{{Hersh}}, \binits{M.J.}},
\bauthor{\bsnm{{Newman}}, \binits{M.D.}},
\bauthor{\bsnm{{Thomas}}, \binits{D.L.}},
\bauthor{\bsnm{{Card}}, \binits{G.L.}},
\bauthor{\bsnm{{Elmore}}, \binits{D.F.}}:
\byear{2003},
\bctitle{{COR1 inner coronagraph for STEREO-SECCHI}}.
In: \beditor{\bsnm{{Keil}}, \binits{S.L.}},
\beditor{\bsnm{{Avakyan}}, \binits{S.V.}} (eds.)
\bbtitle{Innovative Telescopes and Instrumentation for Solar Astrophysics},
\bsertitle{\procspie}
\bseriesno{4853},
\bfpage{1}.
\adsurl{2003SPIE.4853....1T}.
\end{bchapter}
\endbibitem

\bibitem[\protect\citeauthoryear{{T{\"o}r{\"o}k} and {Kliem}}{2003}]{Tor03}
\begin{barticle}
\bauthor{\bsnm{{T{\"o}r{\"o}k}}, \binits{T.}},
\bauthor{\bsnm{{Kliem}}, \binits{B.}}:
\byear{2003},
\batitle{{The evolution of twisting coronal magnetic flux tubes}}.
\bjtitle{\aap}
\bvolume{406},
\bfpage{1043}.
\doiurl{10.1051/0004-6361:20030692}.
\adsurl{2003A\%26A...406.1043T}.
\end{barticle}
\endbibitem

\bibitem[\protect\citeauthoryear{{Tziotziou}, {Georgoulis}, and
  {Liu}}{2013}]{Tzi13}
\begin{barticle}
\bauthor{\bsnm{{Tziotziou}}, \binits{K.}},
\bauthor{\bsnm{{Georgoulis}}, \binits{M.K.}},
\bauthor{\bsnm{{Liu}}, \binits{Y.}}:
\byear{2013},
\batitle{{Interpreting Eruptive Behavior in NOAA AR 11158 via the Region's
  Magnetic Energy and Relative-helicity Budgets}}.
\bjtitle{\apj}
\bvolume{772},
\bfpage{115}.
\doiurl{10.1088/0004-637X/772/2/115}.
\adsurl{2013ApJ...772..115T}.
\end{barticle}
\endbibitem

\bibitem[\protect\citeauthoryear{{Vourlidas} \textit{et~al.}}{2011}]{Vou11}
\begin{barticle}
\bauthor{\bsnm{{Vourlidas}}, \binits{A.}},
\bauthor{\bsnm{{Colaninno}}, \binits{R.}},
\bauthor{\bsnm{{Nieves-Chinchilla}}, \binits{T.}},
\bauthor{\bsnm{{Stenborg}}, \binits{G.}}:
\byear{2011},
\batitle{{The First Observation of a Rapidly Rotating Coronal Mass Ejection in
  the Middle Corona}}.
\bjtitle{\apjl}
\bvolume{733},
\bfpage{L23}.
\doiurl{10.1088/2041-8205/733/2/L23}.
\adsurl{2011ApJ...733L..23V}.
\end{barticle}
\endbibitem

\bibitem[\protect\citeauthoryear{{Zhang} and {Dere}}{2006}]{Zha06}
\begin{barticle}
\bauthor{\bsnm{{Zhang}}, \binits{J.}},
\bauthor{\bsnm{{Dere}}, \binits{K.P.}}:
\byear{2006},
\batitle{{A Statistical Study of Main and Residual Accelerations of Coronal
  Mass Ejections}}.
\bjtitle{\apj}
\bvolume{649},
\bfpage{1100}.
\doiurl{10.1086/506903}.
\adsurl{2006ApJ...649.1100Z}.
\end{barticle}
\endbibitem

\end{thebibliography}

\end{article}
\end{document}